\documentclass[aps,prl,twocolumn,groupedaddress,superscriptaddress,showpacs]{revtex4-1}

\usepackage{graphicx,psfrag,color,hyperref}
\usepackage{amssymb,latexsym,mathrsfs,amsmath,amsthm}
\usepackage[utf8]{inputenc}

\begin{document}

\def\be{\begin{equation}}
\def\ee{\end{equation}}
\def\bea{\begin{eqnarray}}
\def\eea{\end{eqnarray}}

\title{Fisher information from stochastic quantum measurements}

\author{Matthias M. M\"uller}
\affiliation{\mbox{LENS and QSTAR, via Carrara 1, I-50019 Sesto Fiorentino, Italy}}
\affiliation{\mbox{Physics Department, University of Florence}, via G. Sansone 1, I-50019 Sesto Fiorentino, Italy}

\author{Stefano Gherardini}
\affiliation{\mbox{LENS and QSTAR, via Carrara 1, I-50019 Sesto Fiorentino, Italy}}
\affiliation{\mbox{Physics Department, University of Florence}, via G. Sansone 1, I-50019 Sesto Fiorentino, Italy}
\affiliation{\mbox{INFN and Department of Information Engineering, University of Florence,} via S. Marta 3, I-50139 Florence, Italy}

\author{Augusto Smerzi}
\affiliation{\mbox{LENS and QSTAR, via Carrara 1, I-50019 Sesto Fiorentino, Italy}}
\affiliation{\mbox{INO-CNR}, Largo E. Fermi 2, I-50125 Florence, Italy}

\author{Filippo Caruso}
\affiliation{\mbox{LENS and QSTAR, via Carrara 1, I-50019 Sesto Fiorentino, Italy}}
\affiliation{\mbox{Physics Department, University of Florence}, via G. Sansone 1, I-50019 Sesto Fiorentino, Italy}

\begin{abstract}
The unavoidable interaction between a quantum system and the external noisy environment can be mimicked by a sequence of stochastic measurements whose outcomes are neglected. Here we investigate how this stochasticity is reflected in the survival probability to find the system in a given Hilbert subspace at the end of the dynamical evolution. In particular, we analytically study the distinguishability of two different stochastic measurement sequences in terms of a new Fisher information measure depending on the variation of a function, instead of a finite set of parameters. We find a novel characterization of Zeno phenomena as the physical result of the random observation of the quantum system, linked to the sensitivity of the survival probability with respect to an arbitrary small perturbation of the measurement stochasticity. Finally, the implications on the Cramér-Rao bound are discussed, together with a numerical example. These results are expected to provide promising applications in quantum metrology towards future, more robust, noise-based quantum sensing devices.
\end{abstract}

\date{\today}

\maketitle

\paragraph{Introduction.---}

Interaction with an external environment lies at the heart of the dynamical characterization of an open quantum system. No quantum system, indeed, is completely isolated, and it is always characterized by a non-unitary evolution of its state \cite{Petruccione1,Caruso14}. Local coupling effects of the system with the outside can be described as projection events due to the action of one (or more) measurement operators \cite{Zoller1}, along the lines of formalism of the quantum jump trajectories \cite{Plenio1} for open quantum systems. Moreover, as one may expect, these trajectories resulting by the dissipative action of the environment are intrinsically stochastic processes, since any interaction shall occur at irregular time intervals, in general without any a-priori predictability. In this context, it becomes important to investigate the distinguishability of quantum states \cite{Wootters1,Braunstein1} of a randomly perturbed quantum system when also the characterization of the stochasticity rate affecting the system is taken into account.

Recently, it has been shown that statistical indistinguishability of neighboring quantum states can be inferred in terms of measurement results by the evaluation of the system dynamical behaviour in the Zeno regime \cite{Smerzi1}. Indeed, quantum Zeno (QZ) phenomena can be obtained by observing a quantum system by a frequent enough sequence of measurements bringing the system back to its initial state \cite{Misra1,Pascazio1}. As a matter of fact, an unstable quantum system, if observed continuously, will never decay, and its evolution remains frozen. As main application, the Zeno effect has been theoretically exploited to preserve coherent dynamics in a specific subspace of the Hilbert space, by the creation of decoherence-free regions \cite{Paz_Silva1,Maniscalco1}, and it has been experimentally confirmed first with a rubidium Bose–Einstein condensate in a five-level Hilbert space \cite{Schafer1} and later in a multi-level Rydberg state structure \cite{Signoles1}.

The relation between Zeno phenomena and stochastically measured quantum systems has been recently proposed \cite{Shushin1}, and, particularly, in Ref. \cite{Gherardini1} it has been proved that the probability to find the quantum system in the projected state at an arbitrary time instant (survival probability) takes a large deviation (LD) form in the limit of large number of projection events. The LD theory deals with the exponential decay of probabilities of large fluctuations in random systems \cite{Ellis1,Touchette1,Dembo1}. Then, its extension to the quantum case has allowed some of us to analyze the spreading of the system quantum trajectories outside the measurement subspace, and to find the conditions when the ergodic hypothesis for a randomly perturbed quantum system can be effectively verified \cite{Gherardini2}.

In this Letter, we introduce and characterize a novel measure for the state distinguishability of a quantum stochastic process resulting by random sequence of repeated measurements. A key role is then played by the Zeno dynamics, whereby the largest interval such that two quantum states remain indistinguishable is usually denoted as the quantum Zeno time. This latter quantity can be written in terms of the Fisher information (FI) related to the conditional probability that the system state, after a free evolution, is projected into the Zeno subspace \cite{Smerzi1}. A FI measure has been recently introduced to investigate the realizability of quantum Zeno dynamics, when non-Markovian noise is also included \cite{Zhang1}, but, as in \cite{Smerzi1}, the small parameter of the theory is the constant time interval between two consecutive measurements. Conversely, within the formalism of stochastic quantum measurements, here we introduce a Fisher Information Operator, for which the dynamical small parameters are defined by the statistical moments of the stochastic noise acting on the quantum system.

\paragraph{Stochastic quantum measurements.---}

Let us consider the time evolution of a quantum mechanical system subject to a sequence of $m$ measurements. The measurements are spaced by random time intervals $\mu_{j}$, $j = 1,\ldots,m$, sampled by the probability density function $p(\mu)$. The system evolution between the instantaneous measurements is given by the equation $\dot{\rho} = \mathcal{L}_{\mu}[\rho]$, where $\rho$ is the density operator describing the quantum state, and $\mathcal{L}_{\mu}[\circ]$ is the Liouvillian operator. The initial state $\rho_0 \equiv \rho(t=0)$ can be mixed or pure. The survival probability after $m$ measurements is $\mathcal{P}(\{\mu_{j}\}) = \prod_{j = 1}^{m}q(\mu_{j})$, where $q(\mu_{j})$ is the survival probability after each measurement and is defined in terms of the measurement operator and the system time evolution, while the function $q$ does not depend itself on $j$. This assumption does hold, for instance, in the case of the projection on a subspace given by the projector $\Pi$ and small time intervals leading to quantum Zeno dynamics~\cite{Smerzi1}. Let us point out that $\mathcal{P}$ is itself a random variable as it depends on the realization of the random time intervals $\{\mu_j\}$ between the single measurements. Then, by means of large deviation theory, it has recently been demonstrated that the survival probability $\mathcal{P}(\{\mu_{j}\})$ will converge to its most probable value $\mathcal{P}^\star = \exp\left\{m\int_{\mu}p(\mu)\ln(q(\mu))d\mu\right\}$, for a large number $m$ of measurements~\cite{Gherardini1}.

Here, we want to investigate the sensitivity of $\mathcal{P}^\star$ with respect to a perturbation $\delta p(\mu)$ of the underlying probability density function $p(\mu)$. Indeed, this perturbation will induce a change of $\mathcal{P}^\star$ by the quantity
\begin{equation}
 \delta\mathcal{P}^\star=m\mathcal{P}^\star\int_\mu \delta p(\mu)\ln q(\mu) d\mu,
\end{equation}
corresponding to the functional derivative $
  \frac{\delta\mathcal{P}^\star}{\delta p}(\cdot)=m\mathcal{P}^\star\int_\mu (\cdot)\ln q(\mu) d\mu$.
Note that formally it is an element of the dual space of the tangent space of the probability density functions, thus a linear mapping from the admissible changes $\delta p(\mu)$ to a real number $\delta \mathcal{P}^\star$. We can express this fact by the ket notation $\langle \cdot |$, such that $\Big\langle\frac{\delta\mathcal{P}^\star}{\delta p}\Big|=m \mathcal{P}^\star \langle\ln q |$ and for two arbitrary functions $f$ and $g$ the applicaton of a bra to a ket reads $\langle f|g\rangle=\int_\mu f(\mu) g(\mu) d\mu$.
If the projective measurements are frequent enough, the system evolution is effectively limited to the subspace given by the measurement projector $\Pi$, such that in the limit of infinite measurement frequency the survival probability given by its most probable value $\mathcal{P}^\star$ converges to one. The small deviation from this ideal scenario can be approximated by
\begin{equation}\label{eq:Zeno}
\mathcal{P}^\star\approx 1 + \Big\langle\frac{\delta\mathcal{P}^\star}{\delta p}\Big| p\Big\rangle,
\end{equation}
i.e. the quality of Zeno confinement is determined by the sensitivity of the survival probability $\mathcal{P}^\star$ with respect to a perturbation $\delta p(\mu)$.

\begin{figure}[t!]
 \centering
 \includegraphics[width=0.48\textwidth]{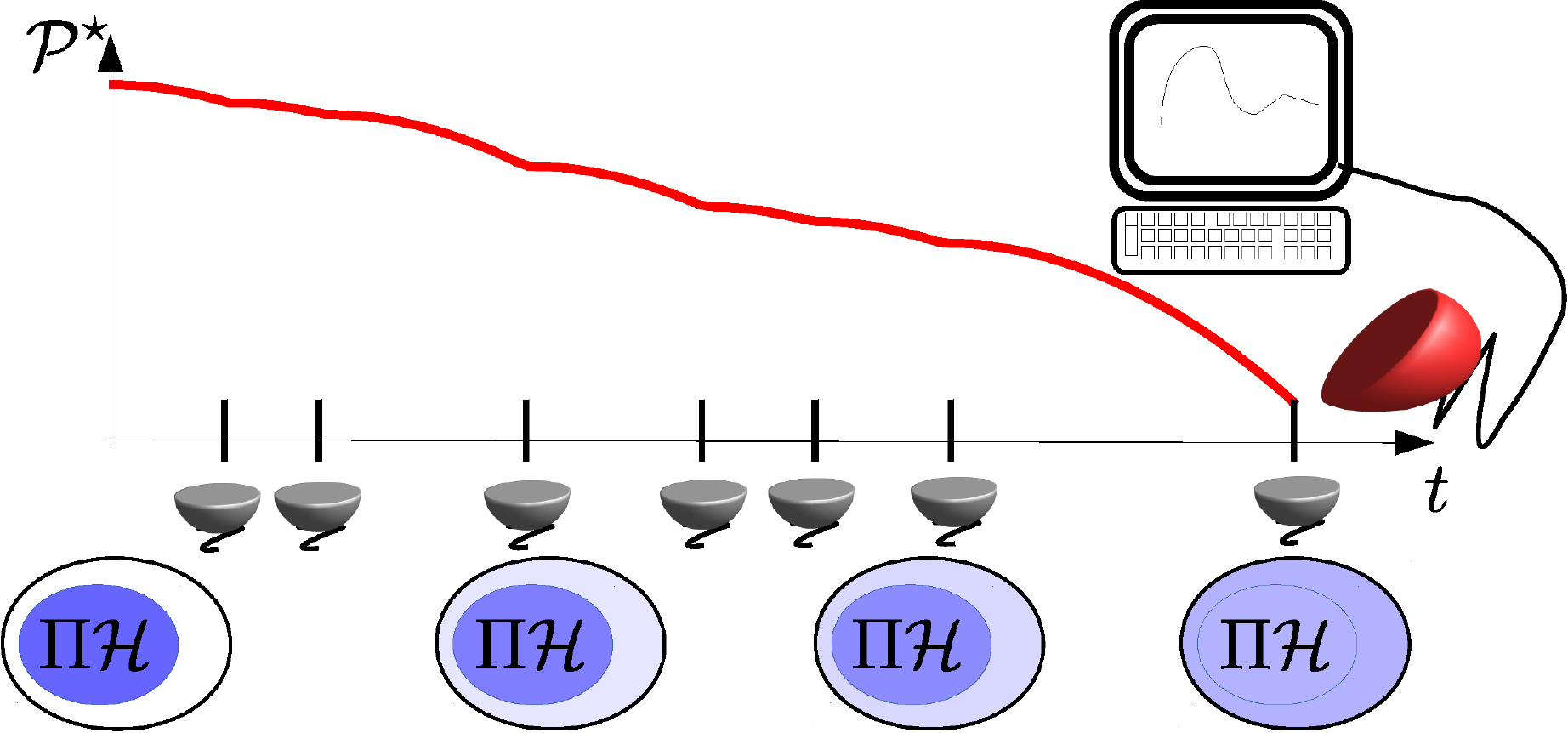}
\caption{Decay of the survival probability $\mathcal{P}^\star$ for a quantum system to remain in an Hilbert subspace when subjected to a stochastic sequence of measurements. As the time goes on, the population slowly leaks out of the subspace ($\Pi \mathcal{H}$, where $\Pi$ is a projector and $\mathcal{H}$ is the full Hilbert space) as illustrated by the blue shades in the lower panel. After each measurement $\mathcal{P}^\star$ evolves quadratically in time. Only the final survival probability is registered by a (red) detector.}
\label{fig1}
\end{figure}

\paragraph{Fisher Information.---}
This sensitivity of the survival probability is very closely linked to the corresponding Fisher information, i.e. the information on $p(\mu)$ that can be extracted by a statistical measurement of $\mathcal{P}^\star$. When dealing with a single estimation parameter $\theta$ and possible measurement results $\eta$, the Fisher information is defined as
$F(\theta)= \int_{\eta}\frac{1}{P(\eta|\theta)}\left(\frac{\partial P(\eta|\theta)}
{\partial\theta}\right)^{2}d\eta$,
which in the case of a binary event, i.e. $\eta\in\{\text{yes},\text{no}\}$, reduces to
\be
F(\theta)= \frac{1}{P(\text{yes}|\theta)(1-P(\text{yes}|\theta))}\left(\frac{\partial P(\text{yes}|\theta)}
{\partial\theta}\right)^{2}.
\ee
Now let us consider the case where we perform $m$ projective measurements on the quantum system, and we keep only the result of the last measurement. As shown in Fig.~\ref{fig1}, we measure survival or not, hence one of two possible events with respective probabilities $\mathcal{P}^\star$ and $1-\mathcal{P}^\star$. Given two different probability density functions characterized by their statistical moments, we can ask how the distinguishability among them can be properly measured. Thus, instead of a single parameter $\theta$, the probability depends on a function $p(\mu)$. We approach this problem by generalizing the Fisher information matrix (FIM)
$
F_{ij}(\underline{\theta})= \frac{1}{P(\text{yes}|\underline{\theta})(1-P(\text{yes}|\underline{\theta}))}\left(\frac{\partial P(\text{yes}|\underline{\theta})}
{\partial\theta_i}\right) \left(\frac{\partial P(\text{yes}|\underline{\theta})}
{\partial\theta_j}\right)\,,
$
depending on the vector $\underline{\theta}=(\theta_1,\theta_2,\dots)$, to a Fisher Information Operator (FIO) involving the functional derivatives of $\mathcal{P}^\star$, as follows
\begin{equation}
 F(p)=m^2\frac{\mathcal{P}^\star}{1-\mathcal{P}^\star}|\ln q\rangle\langle \ln q|\,.
\end{equation}
Note that also $\mathcal{P}^\star$ depends on $m$, so that in the Zeno limit we get a linear scaling of the Fisher information with $m$, i.e. $F(p)\approx m|\ln q\rangle\langle \ln q|/|\langle\ln q|p\rangle|$. Moreover, since our binary measurement outcomes determine just $\mathcal{P}^\star$ and not its distribution, the FIO is a rank one operator.
As a consequence, it is characterized by the single eigenvector $|v\rangle=|\ln q\rangle$ corresponding to the non-zero eigenvalue
\begin{equation}
 F_v=m^2\frac{\mathcal{P}^\star}{1-\mathcal{P}^\star}\Vert \ln q\Vert^{2} \; ,
\end{equation}
with $\Vert \cdot \Vert$ being the $L_2$-norm, namely $\Vert \ln q\Vert^{2} = \int_{\mu}|\ln q(\mu)|^{2}d\mu$.

It can be desirable to express the FIO in a certain basis, thus, transforming it into a FIM. For a given basis $\{f_i\}$ the elements of the FIM read
$ F_{ij}=m^2\frac{\mathcal{P}^\star}{1-\mathcal{P}^\star}\langle f_i|\ln q\rangle\langle \ln q|f_j\rangle$.
In particular, we might be interested in expressing the FIO in terms of the statistical moments
$ \chi_k=\int_{\mu}p(\mu)\mu^k d\mu$
of the probability density function $p(\mu)$. The respective basis functions, then, are given by
$f_k(\mu)=2 \frac{(-1)^k}{k!}\frac{\partial^k}{\partial\mu^k}\delta_{Dirac}(\mu)$, where $\delta_{Dirac}(\mu)$ is the Dirac delta function.
By means of a Taylor expansion around zero we can express $\ln q(\mu)$ by
$
\ln(q(\mu)) = \sum_{k = 1}^{\infty}\left.\frac{\partial^{k}\ln(q(\mu))}{\partial\mu^{k}}\right|_{\mu = 0}\frac{\mu^k}{k!},
$
and, by defining ${\beta_{k}\equiv\left.\frac{\partial^{k}\ln(q(\mu))}{\partial\mu^{k}}\right|_{\mu =0}}$, we obtain
\be\label{eq:Pstar-expansion}
\mathcal{P}^{\star} = \exp\left\{m\sum_{k = 1}^{\infty}\frac{\beta_{k}\chi_{k}}{k!}\right\},
\ee
as well as $\langle f_i|\ln q\rangle = \frac{\beta_i}{i!}$. The resulting FIM reads
\be\label{eq:Fisher_element}
F_{ij}(\underline{\chi}) = m^{2}\frac{\mathcal{P}^{\star}}{(1-\mathcal{P}^{\star})}\frac{\beta_{i}\beta_{j}}{i!j!},
\ee
where the (infinite dimensional) vector $\underline{\chi}$ contains the statistical moments $ \chi_k$ of $p(\mu)$.
Notice that this result is compatible with the standard definition of the FIM
$ F_{ij}(\underline{\chi}) =\frac{1}{\mathcal{P}^{\star}(1-\mathcal{P}^{\star})}\frac{\partial \mathcal{P}^{\star}}{\partial\chi_i}\frac{\partial \mathcal{P}^{\star}}{\partial\chi_j}$,
since $\displaystyle{\frac{\partial \mathcal{P}^{\star}}{\partial\chi_{h}} = m\mathcal{P}^{\star}\frac{\beta_{h}}{h!}}$.
As observed for the FIO, the rank of the FIM is equal to one~\cite{note}. This implies that, in principle, we can distinguish two probability density functions that differ by a single statistical moment or a linear combination of them. The highest sensitivity of such a distinguishability problem is found for a difference of the statistical moments along the (single) eigenvector $\underline{v}$ corresponding to the non-zero eigenvalue $\tilde{F}_{v}$ of the FIM. This eigenvalue is given by
\be\label{eq:eigenvalue}
\tilde{F}_{v} = m^{2}\frac{\mathcal{P}^{\star}}{1-\mathcal{P}^{\star}}\sum_k\left(\frac{\beta_k}{k!}\right)^2.
\ee
Since the basis functions $\{f_k\}$ were not normalized, it is different from the eigenvalue $F_v$ of the FIO. The $i-$th element of the (non-normalized) eigenvector $\underline{v}$ results to be $\underline{v}_{i} = \beta_{i}/i!$. The most probable value $\mathcal{P}^\star$, thus, can be expressed as a function of $F_{v}$ ($\tilde{F}_v$) and $|v\rangle$ ($\underline{v}$), as follows:
\be\label{eq:relation1}
\mathcal{P}^\star = \frac{F_{v}}{F_{v} + m^{2}\|v\|^{2}}=\frac{\tilde{F}_{v}}{\tilde{F}_{v} + m^{2}\|\underline{v}\|^{2}}
\ee
or equivalently $\mathcal{P}^\star = \exp\{m\langle\underline{v},\underline{\chi}\rangle\}$, where the functions $\|\underline{v}\| = \sqrt{\sum_{k}(v_{k})^{2}}$ and $\langle\underline{v},\underline{\chi}\rangle = \sum_{k}v_{k}\chi_{k}$ are defined, respectively, as
the Euclidian norm and the scalar product. The eigenvector depends only on the system properties ($\underline{\beta}$), while the eigenvalue depends on both the system ($\underline{\beta}$) and the probability density function $p(\mu)$ ($\underline{\chi}$). As a matter of fact, by taking $p(\mu)$ such that its statistical moments $\underline{\chi} = \underline{v}$, the eigenvalue $\tilde{F}_{v}$ of the FIM can be written as a function of only $\mathcal{P}^\star$:
$\tilde{F}_{v} = - m\frac{\mathcal{P}^\star }{1 - \mathcal{P}^\star}\mathcal{I}(\mathcal{P}^\star)$,
where $\mathcal{I}(\mathcal{P}^\star)=-\ln\mathcal{P}^\star$ is the self-information related to the event $\mathcal{P}^\star$.

\paragraph{Zeno-Regime and Cramér-Rao bound.---}
The confinement error $1-\mathcal{P}^\star$ in the Zeno regime can now be expressed in terms of the FIO. Namely, analogously to Eq.~\eqref{eq:Zeno}, we find $1 - \mathcal{P}^\star\propto m \langle v| p \rangle$, where $\langle v| p\rangle$ is the scalar product $\int_{\mu}p(\mu)v(\mu)d\mu=\int_{\mu}p(\mu)\ln q(\mu)d\mu$. Hence, the confinement error is given by the overlap between the distribution $p(\mu)$ with the eigenvector $|v\rangle$ of the FIO. The standard (non-stochastic) Zeno limit is usually obtained by increasing the number of measurements $m$ in a fixed time interval $T$, with the survival probability converging to one~\cite{Smerzi1}. In our case $m$ measurements occur at random times and thus we set $T(m)=m\int_{\mu}p(\mu) d\mu=T=const$, thus, fixing the expectation value of the final time. As a consequence, in order to approach the Zeno limit we need the condition
$
 \lim_{\substack{m\rightarrow\infty\\ T(m)=T}} m \int_{\mu}p(\mu)\ln q(\mu)d\mu\stackrel{!}{=}0
$. In other terms, increasing the number of interactions $m$, the average value of $\ln q(\mu)\approx 1-q(\mu)$ has to decrease at least as $1/m$. In the case of $p(\mu)=\delta(\mu-\tilde{\mu})$ we recover the condition $\lim_{m\rightarrow\infty}m(1-q(\tilde{\mu}))\stackrel{!}{=}0$, which is usually fulfilled when the $m$ single intervals are set to $\tilde{\mu}=T/m$~\cite{Smerzi1}. This is closely linked to a distinguishability problem, i.e. the effect of the deviation of $p(\mu)$ from $\delta(\mu)$ (corresponding to infinitely many measurements) on the decrease of the survival probability and the minimum deviation that can be detected.

This problem can be formulated in terms of the Cramér-Rao bound. Let us consider a distortion in the function space $\delta p^{(f)}(\mu)=\delta c\, f(\mu)$, where $\delta c$ is the parametrization of a small perturbation along the direction $f(\mu)$ in the tangent space of the probability density functions. The Fisher information for estimating the parameter $c$ is given by $F_f=\langle f|F(p)|f\rangle$. The statistical moments corresponding to $f(\mu)$ are $\xi_k^{(f)}=\int_{\mu}f(\mu)\mu^{k}d\mu$. The FIM allows us to express the Fisher information for estimating the parameter $c$ in terms of the statistical moments as
$
F_{\underline{\xi}^{(f)}}=\sum_{i,j}\xi_i^{(f)} F_{ij} \xi_j^{(f)}= m^2\frac{\mathcal{P}^\star}{1-\mathcal{P}^\star}\left(\sum_k \frac{\beta_k\xi_k^{(f)}}{k!}\right)^{2},
$
leading to the Cramér-Rao bound
\begin{eqnarray}\label{eq:CRB-FIM}
 \delta c\geq \frac{1}{\sqrt{F_{\underline{\xi}^{(f)}}}}=\frac{\sqrt{1-\mathcal{P}^\star}}{m\sqrt{\mathcal{P}^\star}\sum_k \frac{\beta_k\xi_k^{(f)}}{k!} }.
\end{eqnarray}
Alternatively, the Cramér-Rao bound can be also directly computed from $\delta p^{(f)}(\mu)$ as
\begin{eqnarray}\label{eq:CRB-FD}
 \delta c\geq \frac{1}{\sqrt{F_f}}=\frac{\sqrt{1-\mathcal{P}^\star}}{m\sqrt{\mathcal{P}^\star}\int_{\mu}f(\mu)\ln q(\mu)d\mu},
\end{eqnarray}
which is equivalent to Eq.~\eqref{eq:CRB-FIM} because of $\int_{\mu}f(\mu)\ln q(\mu)d\mu=\sum_k\frac{\beta_k\xi_k^{(f)}}{k!}$.

\paragraph{Example.---}
In the case of coherent evolution with Hamiltonian $H$, i.e. $\dot{\rho} = -i[H,\rho]$, the single measurement quantum survival probability $q(\mu_{j})$ is an even function, and, thus, all the odd coefficients $\beta_{k}$, $k = 1,3,\ldots$, of the Taylor expansion of $\ln(q(\mu))$ are identically equal to zero. In the Zeno regime (for small time intervals) one has $q(\mu_{j}) \approx 1-\Delta^2 H_{\Pi} \mu_j^2$, with $H_{\Pi}=H-\Pi H\Pi$, and the variance being calculated with respect to the initial state $\rho_0$. In other terms, the leakage is given by the terms in the Hamiltonian connecting the measurement subspace with the rest of the Hilbert space. The corresponding survival probability after $m$ measurements can be naturally approximated to the second order of the time interval length by the contribution of $\chi_2$, i.e. $\mathcal{P}^{\star}=1 - m\Delta^{2}H_{\Pi}\chi_2$. This allows us to express the survival probability in terms of the relevant element $F_{22}=\frac{m\Delta^{2}H_{\Pi}}{\chi_2}$ of the FIM:
\be\label{eq:P-expansion-chi2}
\mathcal{P}^{\star} = 1 - F_{22}\chi_{2}^{2},
\ee
generalizing, thus, the result obtained for equally time-distributed sequence of projective measurements \cite{Smerzi1}.
As a consequence of Eq.~\eqref{eq:P-expansion-chi2}, we can straightforwardly derive the Cramér-Rao lower bound for estimating the second moment $\chi_2$ from $\mathcal{P}^{\star}$, i.e. $\displaystyle{\Delta\chi_{2} \geq \frac{1}{\sqrt{F_{22}}} = \frac{\sqrt{\chi_2}}{\sqrt{m}\Delta H_{\Pi}}}$, which provides a natural condition for the statistical moment indistinguishability in terms of the ratio $\chi_{2}/\Delta\chi_{2}$.

These analytical results can be applied to distinguish two probability density functions $p(\mu)$, modeling the random interaction of a quantum system with the outside, and differing for a linear combination of their statistical moments. Let us consider, thus, an uniform probability distribution function $p(\mu)$ on the interval $[\mu_{1},\mu_{2}]$, i.e.  $p(\mu)= (\Theta(\mu-\mu_1)-\Theta(\mu-\mu_2))/(\mu_2-\mu_1)$, where $\Theta$ is the Heaviside function.
\begin{figure}[t!]
 \centering
 \includegraphics[width=0.48\textwidth]{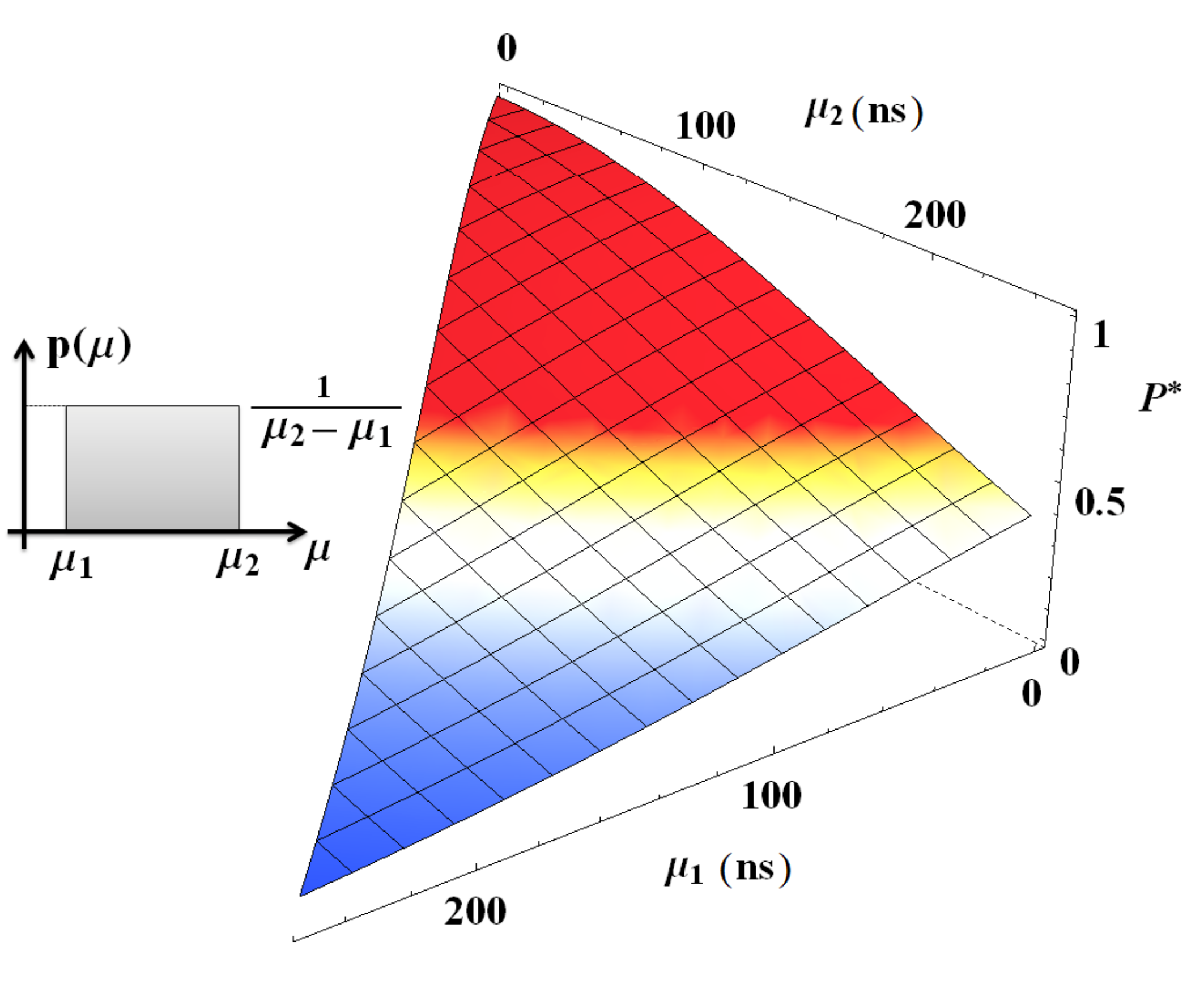}
\caption{Most probable value $\mathcal{P}^{\star}$ as a function of $\mu_{1}$ and $\mu_{2}$. The colour scale (from blue to red) of the surface refers to the values assumed by the non-zero eigenvalue $\tilde{F}_{v}$ of the corresponding FIM, normalized to the Euclidian norm of the eigenvector. We consider a local spin Hamiltonian with $N = 9$ spins, $m = 5000$, $\omega = 2\pi\times 5$ kHz, and $\Delta^2 H_{\Pi}$ set to its upper bound $\omega^{2}N$ holding for product states $|\psi_0\rangle=|00...0\rangle$. Inset: uniform probability distribution $p(\mu)$ in the range $[\mu_1,\mu_2]$.}
\label{fig2}
\end{figure}
The second moment of this distribution is given by
$
\chi_2=\int_{\mu_1}^{\mu_2}p(\mu)\mu^2 d\mu=\frac{\mu_2^3-\mu_1^3}{3(\mu_2-\mu_1)}.
$
If we only change $\mu_2$, the partial derivative of the second moment is
$
\frac{\partial \chi_2}{\partial \mu_2}=\frac{-3\mu_2^{2}\mu_1+ 2\mu_2^3+\mu_1^3}{3(\mu_2-\mu_1)^2}$.
The probability density function, instead, changes by $\delta p^{(f)}(\mu)=\delta c\, f(\mu)$ with $f(\mu)=\frac{\delta_{Dirac}(\mu-\mu_2)-p(\mu)}{\mu_2-\mu_1}$, leading to the Fisher information
\bea\label{eq:Fisher-uniform-mu2}
F_f=
m^2\frac{\mathcal{P}^{\star}}{1-\mathcal{P}^{\star}}\left\langle f|  \ln q(\mu)\right\rangle^2
\approx m\Delta^{2}H_{\Pi} \frac{(\mu_2^2-\chi_2)^2}{\chi_2(\mu_2-\mu_1)^2} \; . \ \ \ \ \ \
\eea
On the other side, if we consider the sensitivity with respect to a change in $\chi_2$, the Fisher information is given by the respective matrix element $F_{22}\approx m\Delta^2 H/\chi_2$, as in Eq.~(\ref{eq:P-expansion-chi2}). Because of the constraint in the distribution shape variation, the two results differ by the square of the derivative of the second moment with respect to the parameter, due to the derivative chain rule
$\frac{F(f(\mu))}{F(\chi_2) (\frac{\partial}{\partial \mu_2}\chi_2)^2}=1$ -- see SI for more details. Now, let us consider the local spin Hamiltonian
$H=\omega\sum_{n=1}^N \vec{\alpha_n}\cdot\vec{\sigma}$, with $N$ spins and the normalized real coefficient vectors $\vec{\alpha_n}$ for the three Pauli spin matrices $\vec{\sigma}=(\sigma_x,\sigma_y,\sigma_z)^T$. We span the Zeno subspace by the initial state, i.e. the projector $\Pi=|\psi_0\rangle\langle\psi_0|$. This limits the variance to $\Delta^2 H_{\Pi} \leq N\omega^2$ for product states and to $\Delta^2 H_{\Pi} \leq \omega^2 N^2$ for entangled ones. This is also a bound to the single measurement Fisher information $F_{sm}$ in the Zeno limit, i.e. $F_{sm}=4\Delta^2 H_{\Pi}$~\cite{Smerzi1,Pezze}. In our case we have an additional factor depending on $p(\mu)$ as given by Eq.~(\ref{eq:Fisher-uniform-mu2}). We can saturate the bound for product states with $|\psi_0\rangle=|00...0\rangle$ and $\vec{\alpha_n}=(1,0,0)$, as well as the bound for entangled states is with the GHZ state $(|0..00\rangle-i|1..1\rangle)/\sqrt{2}$ and $\vec{\alpha_n}=(0,0,1)$.
\begin{figure}[t!]
 \centering
 \includegraphics[width=0.48\textwidth]{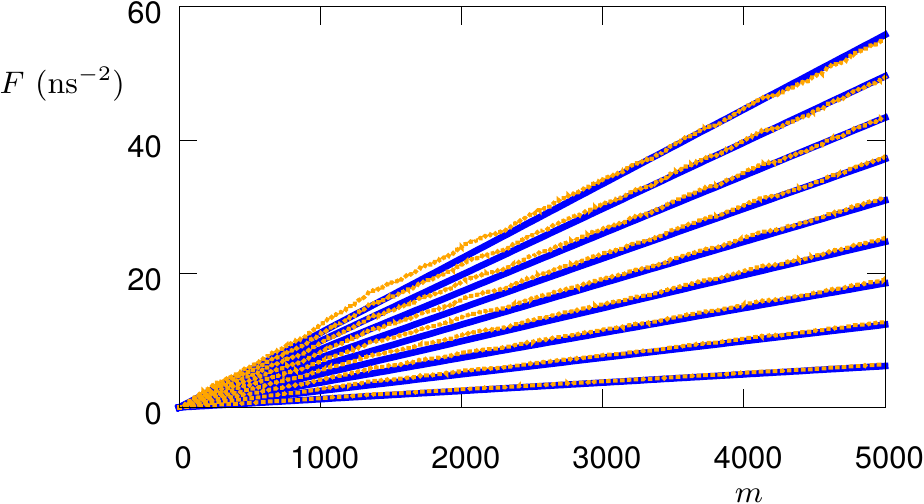}
\caption{Fisher information for determining the upper bound $\mu_2$ of the uniform probability distribution $p(\mu)$ (simulation orange dots) saturating the bound (blue lines) for product states and exhibiting linear scaling with the number of measurements $m$ and the number of spins $N=1,..,9$ (from bottom to top). The parameters are $\omega=2\pi\times 5\,$kHz, $\mu_1=10\,$ns, $\mu_2=60\,$ns.}
\label{fig:uniform1}
\end{figure}
In Fig. \ref{fig2}, we show the behaviour of $\mathcal{P}^\star$ as a function of $\mu_1$ and $\mu_2$, and the corresponding value of $\tilde{F}_{v}$. The larger is the latter value, the higher is the sensitivity to distinguish finite perturbations $\delta p(\mu)$ by measuring $\mathcal{P}^{\star}$ after $m$ measurements. It occurs when approaching the Zeno regime, with $\mathcal{P}^{\star}$ being close to $1$. Finally, Fig.~\ref{fig:uniform1} illustrates the Fisher information for a variation in $\mu_2$, and its linear scaling with $m$ and $N$ as compared to the theoretical bound for product states. The Fisher information for $m=5000$ is $F(N)\approx N\times6.5\,\mathrm{ns}^{-2}$. This corresponds to a Cramér-Rao bound for the parameter $\mu_2$ of $\delta\mu_2\geq \frac{1}{\sqrt{F(N)}}\approx \frac{1}{\sqrt{N}}\times 0.39\,\mathrm{ns}$. Depending on the number of qubits $N$ the relative error is $0.2-0.7\,\%$.

\paragraph{Conclusions and Outlooks.---}

We have analytically derived the conditions under which two neighboring quantum states, resulting from a sequence of stochastic quantum measurements on the system, can be effectively distinguished. In particular, by exploiting large deviation theory, we have introduced a Fisher Information Operator (FIO), expressed in terms of the statistical moments of the probability density function $p(\mu)$ defining the random nature of the interactions. This has allowed us to properly analyze the sensitivity of the most probable value of the probability for the system to be in the measurement subspace with respect to an arbitrary perturbation $\delta p(\mu)$ of $p(\mu)$, and to determine the corresponding Cramér-Rao bound. Finally, a numerical example is shown, in which a parameter of a uniform probability density function is estimated with high precision. These results, on one side, may trigger a widespread interest for its foundational implications about the nature of quantum Zeno phenomena, and, on the other, may find promising applications for robust quantum information processing in driving the system dynamics along various quantum paths within the whole Hilbert space, and also in the context of quantum metrology.

\

\begin{acknowledgments}
We acknowledge fruitful discussions with S. Ruffo, S. Gupta and F.S. Cataliotti. This work was supported by the Seventh Framework Programme for Research of the European Commission, under the CIG grant QuantumBioTech, and by the Italian Ministry of Education, University and Research (MIUR), under FIRB Grant Agreement No. RBFR10M3SB.
M.M. and S.G. contributed equally to this work.
\end{acknowledgments}

\appendix

\newpage

\section{Supplementary Information}

\subsection{Fisher derivation for the uniform distribution $p(\mu)$}

Let us assume the standard Zeno approximations $q(\mu)\approx 1-\Delta^2H_{\Pi} \mu^2$, $\mathcal{P}^{\star}=1$, $(1-\mathcal{P}^{\star})=m\Delta^2H_{\Pi}\chi_2$, and consider the uniform probability distribution $p(\mu)= (\Theta(\mu-\mu_1)-\Theta(\mu-\mu_2))/(\mu_2-\mu_1)$, where $\Theta$ is the Heaviside function. The second moment of this distribution is given by
\begin{equation*}
\chi_2=\int_{\mu_1}^{\mu_2}p(\mu)\mu^2 d\mu=\frac{\mu_2^3-\mu_1^3}{3(\mu_2-\mu_1)}=\frac{\mu_1^2+\mu_1\mu_2+\mu_2^2}{3}.
\end{equation*}
Accordingly, the $k$-th moment is given by
\begin{equation*}
\chi_k=\frac{\mu_2^{k+1}-\mu_1^{k+1}}{(k+1)(\mu_2-\mu_1)}.
\end{equation*}
If we change $\mu_2$, then the partial derivatives of the statistical moments are
\begin{equation*}
\frac{\partial \chi_k}{\partial \mu_2}=\frac{(k+1)\mu_2^{k}(\mu_2-\mu_1)-(\mu_2^{k+1}-\mu_1^{k+1})}{(k+1)(\mu_2-\mu_1)^2} \; .
\end{equation*}
In particular, for the second moment, we have
\begin{equation*}
\frac{\partial \chi_2}{\partial \mu_2}=\frac{3\mu_2^{2}(\mu_2-\mu_1)-(\mu_2^3-\mu_1^3)}{3(\mu_2-\mu_1)^2}
=\frac{-3\mu_2^{2}\mu_1+ 2\mu_2^3+\mu_1^3}{3(\mu_2-\mu_1)^2}.
\end{equation*}
The probability density function, instead, changes by $\delta p^{(f)}(\mu)=\delta c\, f(\mu)$ with $f(\mu)=\frac{\delta_{Dirac}(\mu-\mu_2)-p(\mu)}{\mu_2-\mu_1}$, leading to the Fisher information
\begin{equation*}
\begin{split}
&F_{f}=\frac{\mathcal{P}^{\star}}{1-\mathcal{P}^{\star}}\left(m\int_{\mu}\frac{\delta_{Dirac}(\mu-\mu_2)-p(\mu)}{\mu_2-\mu_1}  \ln q(\mu)d\mu\right)^2 \\
&\approx \frac{\mathcal{P}^{\star}}{1-\mathcal{P}^{\star}}\left(m\Delta^2 H_{\Pi}\int_{\mu}\frac{\delta_{Dirac}(\mu-\mu_2)-p(\mu)}{\mu_2-\mu_1} \mu^{2}d\mu\right)^2 \\
&\approx m\Delta^2H_{\Pi}\frac{(\mu_2^2-\chi_2)^2}{\chi_2(\mu_2-\mu_1)^2}.
\end{split}
\end{equation*}
Conversely, if we treat $\chi_2$ as the estimation parameter $\theta$, the Fisher information reads
\begin{equation*}
\begin{split}
&F(\chi_2)=\frac{\mathcal{P}^{\star}}{1-\mathcal{P}^{\star}}\left(m\frac{\partial}{\partial\theta}\int_{\mu}p(\mu)\ln q(\mu)d\mu\right)^{2} \\
&\approx\frac{\mathcal{P}^{\star}}{1-\mathcal{P}^{\star}}\left(m\frac{\partial}{\partial\chi_2}(\Delta^2 H_{\Pi}\chi_2)\right)^2 \\ &=\frac{\mathcal{P}^{\star}}{1-\mathcal{P}^{\star}}\left( m\Delta^2 H_{\Pi}\right)^2 \approx m\Delta^2 H_{\Pi}/\chi_2.
\end{split}
\end{equation*}
This is also the respective matrix element of the Fisher information matrix. The two results differ because we put a contraint on the shape of the probability distribution. In fact, at the second order (since we neglect higher moments) they differ by the square of derivative of the second moment as the Fisher information follows a chain rules for the derivative:
\begin{equation*}
\begin{split}
&\frac{F(f(\mu))}{F(\chi_2) (\frac{\partial}{\partial \mu_2}\chi_2)^2}=
\frac{(\mu_2^2-\chi_2)^2}{(\mu_2-\mu_1)^2}\left(\frac{3(\mu_2-\mu_1)^2}{-3\mu_2^{2}\mu_1+ 2\mu_2^3+\mu_1^3}\right)^2 \\
&=\frac{(3(\mu_2-\mu_1)\mu_2^2-(\mu_2^3-\mu_1^3))^2}{9(\mu_2-\mu_1)^4}\left(\frac{3(\mu_2-\mu_1)^2}{-2\mu_2^{2}\mu_1+ \mu_2^3+\mu_1^3}\right)^2 \\
&=\left(\frac{ 3(\mu_2-\mu_1)\mu_2^2-(\mu_2^3-\mu_1^3) }{-3\mu_2^{2}\mu_1+ 2\mu_2^3+\mu_1^3}\right)^2=1 \; .
\end{split}
\end{equation*}


\begin{thebibliography}{99}
\bibitem{note}
In order to demonstrate that the rank of the FIM $F(\underline{\chi})$ is $1$, we simply need to show that the determinant of a generic $2\times 2$ minor of the Fisher matrix is equal to $0$, i.e.
\begin{equation*}
\left(m^{2}\frac{\mathcal{P}^{\star}}{(1-\mathcal{P}^{\star})}\right)^2
\begin{vmatrix}
 \frac{\beta_{i}\beta_{j}}{i!j!} &\frac{\beta_{i}\beta_{j+1}}{i!(j+1)!}\\
 \frac{\beta_{i+1}\beta_{j}}{(i+1)!j!}&\frac{\beta_{i+1}\beta_{j+1}}{(i+1)!(j+1)!}
\end{vmatrix}
=0.
\end{equation*}
\end{thebibliography}
\end{document}